# Measurement of the Average $B$-Hadron Lifetime in $Z^0$ Decays Using Reconstructed Vertices*

* K. Abe,(29) I. Abt,(14) C.J. Ahn,(26) T. Akagi,(27) N.J. Allen,(27) W.W. Ash,(27)† D. Aston,(27) K.G. Baird,(25) C. Baltay,(33) H.R. Band,(32) M.B. Barakat,(33) G. Baranko,(10) O. Bardon,(16) T. Barklow,(27) A.O. Bazarko,(11) R. Ben-David,(33) A.C. Benvenuti,(2) T. Bienz,(27) G.M. Bilei,(22) D. Bisello,(21) G. Blaylock,(7) J.R. Bogart,(27) T. Bolton,(11) G.R. Bower,(27) J.E. Brau,(20) M. Breidenbach,(27) W.M. Bugg,(28) D. Burke,(27) T.H. Burnett,(31) P.N. Burrows,(16) W. Busza,(16) A. Calcaterra,(13) D.O. Caldwell,(6) D. Calloway,(27) B. Camanzi,(12) M. Carpinelli,(23) R. Cassell,(27) R. Castaldi,(23)(a) A. Castro,(21) M. Cavalli-Sforza,(7) E. Church,(31) H.O. Cohn,(28) J.A. Coller,(3) V. Cook,(31) R. Cotton,(4) R.F. Cowan,(16) D.G. Coyne,(7) A. D'Oliveira,(8) C.J.S. Damerell,(24) M. Daoudi,(27) R. De Sangro,(13) P. De Simone,(13) R. Dell'Orso,(23) M. Dima,(9) P.Y.C. Du,(28) R. Dubois,(27) B.I. Eisenstein,(14) R. Elia,(27) D. Falciai,(22) C. Fan,(10) M.J. Fero,(16) R. Frey,(20) K. Furuno,(20) T. Gillman,(24) G. Gladding,(14) S. Gonzalez,(16) G.D. Hallewell,(27) E.L. Hart,(28) Y. Hasegawa,(29) S. Hedges,(4) S.S. Hertzbach,(17) M.D. Hildreth,(27) J. Huber,(20) M.E. Huffer,(27) E.W. Hughes,(27) H. Hwang,(20) Y. Iwasaki,(29) D.J. Jackson,(24) P. Jacques,(25) J. Jaros,(27) A.S. Johnson,(3) J.R. Johnson,(32) R.A. Johnson,(8) T. Junk,(27) R. Kajikawa,(19) M. Kalelkar,(25) I. Karliner,(14) H. Kawahara,(27) H.W. Kendall,(16) Y. Kim,(26) M.E. King,(27) R. King,(27) R.R. Kofler,(17) N.M. Krishna,(10) R.S. Kroeger,(18) J.F. Labs,(27) M. Langston,(20) A. Lath,(16) J.A. Lauber,(10) D.W.G. Leith,(27) X. Liu,(7) M. Loreti,(21) A. Lu,(6) H.L. Lynch,(27) J. Ma,(31) G. Mancinelli,(22) S. Manly,(33) G. Mantovani,(22) T.W. Markiewicz,(27) T. Maruyama,(27) R. Massetti,(22) H. Masuda,(27) E. Mazzucato,(12) A.K. McKemey,(4) B.T. Meadows,(8) R. Messner,(27) P.M. Mockett,(31) K.C. Moffeit,(27) B. Mours,(27) G. Müller,(27) D. Muller,(27) T. Nagamine,(27) U. Nauenberg,(10) H. Neal,(27) M. Nussbaum,(8) Y. Ohnishi,(19) L.S. Osborne,(16) R.S. Panvini,(30) H. Park,(20) T.J. Pavel,(27) I. Peruzzi,(13)(b) M. Piccolo,(13) L. Piemontese,(12) E. Pieroni,(23) K.T. Pitts,(20) R.J. Plano,(25) R. Prepost,(32) C.Y. Prescott,(27) G.D. Punkar,(27) J. Quigley,(16) B.N. Ratcliff,(27) T.W. Reeves,(30) P.E. Rensing,(27) L.S. Rochester,(27) J.E. Rothberg,(31) P.C. Rowson,(11) J.J. Russell,(27) O.H. Saxton,(27) T. Schalk,(7) R.H. Schindler,(27) U. Schneekloth,(16) B.A. Schumm,(15) A. Seiden,(7) S. Sen,(33) V.V. Serbo,(32) M.H. Shaevitz,(11) J.T. Shank,(3) G. Shapiro,(15) S.L. Shapiro,(27) D.J. Sherden,(27) C. Simopoulos,(27) N.B. Sinev,(20) S.R. Smith,(27) J.A. Snyder,(33) P. Stamer,(25) H. Steiner,(15) R. Steiner,(1) M.G. Strauss,(17) D. Su,(27) F. Suekane,(29) A. Sugiyama,(19) S. Suzuki,(19) M. Swartz,(27) A. Szumilo,(31) T. Takahashi,(27) F.E. Taylor,(16) E. Torrence,(16) J.D. Turk,(33) T. Usher,(27) J. Va'vra,(27) C. Vannini,(23) E. Vella,(27) J.P. Venuti,(30) R. Verdier,(16) P.G. Verdini,(23) S.R. Wagner,(27) A.P. Waite,(27) S.J. Watts,(4) A.W. Weidemann,(28) E.R. Weiss,(31) J.S. Whitaker,(3) S.L. White,(28) F.J. Wickens,(24) D.A. Williams,(7) D.C. Williams,(16) S.H. Williams,(27) S. Willocq,(33) R.J. Wilson,(9) W.J. Wisniewski,(5) M. Woods,(27) G.B. Word,(25) J. Wyss,(21) R.K. Yamamoto,(16) J.M. Yamartino,(16) X. Yang,(20) S.J. Yellin,(6) C.C. Young,(27) H. Yuta,(29) G. Zapalac,(32) R.W. Zdarko,(27) C. Zeitlin,(20) Z. Zhang,(16) and J. Zhou,(20)

[1]*Adelphi University, Garden City, New York 11530*
[2]*INFN Sezione di Bologna, I-40126 Bologna, Italy*
[3]*Boston University, Boston, Massachusetts 02215*
[4]*Brunel University, Uxbridge, Middlesex UB8 3PH, United Kingdom*
[5]*California Institute of Technology, Pasadena, California 91125*
[6]*University of California at Santa Barbara, Santa Barbara, California 93106*
[7]*University of California at Santa Cruz, Santa Cruz, California 95064*
[8]*University of Cincinnati, Cincinnati, Ohio 45221*
[9]*Colorado State University, Fort Collins, Colorado 80523*
[10]*University of Colorado, Boulder, Colorado 80309*
[11]*Columbia University, New York, New York 10027*
[12]*INFN Sezione di Ferrara and Università di Ferrara, I-44100 Ferrara, Italy*
[13]*INFN Lab. Nazionali di Frascati, I-00044 Frascati, Italy*
[14]*University of Illinois, Urbana, Illinois 61801*
[15]*Lawrence Berkeley Laboratory, University of California, Berkeley, California 94720*
[16]*Massachusetts Institute of Technology, Cambridge, Massachusetts 02139*
[17]*University of Massachusetts, Amherst, Massachusetts 01003*
[18]*University of Mississippi, University, Mississippi 38677*
[19]*Nagoya University, Chikusa-ku, Nagoya 464 Japan*
[20]*University of Oregon, Eugene, Oregon 97403*
[21]*INFN Sezione di Padova and Università di Padova, I-35100 Padova, Italy*
[22]*INFN Sezione di Perugia and Università di Perugia, I-06100 Perugia, Italy*
[23]*INFN Sezione di Pisa and Università di Pisa, I-56100 Pisa, Italy*
[25]*Rutgers University, Piscataway, New Jersey 08855*
[24]*Rutherford Appleton Laboratory, Chilton, Didcot, Oxon OX11 0QX United Kingdom*
[26]*Sogang University, Seoul, Korea*
[27]*Stanford Linear Accelerator Center, Stanford University, Stanford, California 94309*
[28]*University of Tennessee, Knoxville, Tennessee 37996*
[29]*Tohoku University, Sendai 980 Japan*
[30]*Vanderbilt University, Nashville, Tennessee 37235*
[31]*University of Washington, Seattle, Washington 98195*
[32]*University of Wisconsin, Madison, Wisconsin 53706*
[33]*Yale University, New Haven, Connecticut 06511*
[†]*Deceased*
[a]*Also at the Università di Genova*
[b]*Also at the Università di Perugia*

*Work supported in part by Department of Energy contracts: DE–FG02–91ER40676 (BU), DE–FG03–92ER40701 (CIT), DE–FG03–91ER40618 (UCSB), DE–FG03–92ER40689 (UCSC), DE–FG03–93ER40788 (CSU), DE–FG02–91ER40672 (Colorado), DE–FG02–91ER40677 (Illinois), DE–AC03–76SF00098 (LBL), DE–FG02–92ER40715 (Massachusetts), DE–AC02–76ER03069 (MIT), DE–FG06–85ER40224 (Oregon), DE–AC03–76SF00515 (SLAC), DE–FG05–91ER40627 (Tennessee), DE–AC02–76ER00881 (Wisconsin), DE-FG02–92ER40704 (Yale); National Science Foundation grants: PHY–91–13428 (UCSC), PHY–89–21320 (Columbia), PHY–92–04239 (Cincinnati), PHY–88–17930 (Rutgers), PHY–88–19316 (Vanderbilt), PHY–92–03212 (Washington); the UK Science and Engineering Research Council (Brunel and RAL); the Istituto Nazionale di Fisica Nucleare of Italy (Bologna, Ferrara, Frascati, Pisa, Padova, Perugia); and the Japan-US Cooperative Research Project on High Energy Physics (Nagoya, Tohoku).

# Abstract

We report a measurement of the average $B$-hadron lifetime using data collected with the SLD detector at the SLC in 1993. An inclusive analysis selected three-dimensional vertices with $B$-hadron lifetime information in a sample of 50k $Z^0$ decays. A lifetime of $1.564 \pm 0.030\,(stat) \pm 0.037\,(syst)$ ps was extracted from the decay length distribution of these vertices using a binned maximum-likelihood method.

Measurements of the $B$-hadron lifetime $\tau_B$ are useful in exploring the physics of $b$-quarks, particularly in determining the weak couplings of the $b$ to lighter quarks. Precise measurements of the average value of $\tau_B$ remain interesting in view of the significant variation in the world average over the past few years (see Refs. [1–3]). Previous determinations of $\tau_B$ have relied on either tight lepton momentum cuts [4] or stringent vertex constraints [5] to isolate $B$ hadron decay vertices. In the method presented here, all decay modes are used with high efficiency.

This letter presents a measurement of $\tau_B$ based on inclusive three-dimensional reconstruction of secondary vertices in $Z^0 \to b\bar{b}$ events. The data for this analysis were collected at the SLAC Linear Collider (SLC) with the SLD experiment in 1993. The analysis uses a topological technique to select vertices with lifetime information and relies on Monte Carlo (MC) modeling to extract $\tau_B$, using a maximum-likelihood method. This yielded a precise measurement of $\tau_B$, even with a relatively small data sample.

During the 1993 run, SLD recorded 1.8 $pb^{-1}$ of $e^+e^-$ annihilation data at a center-of-mass energy of $91.26 \pm 0.02$ GeV. Charged particle tracking was provided by the central drift chamber (CDC) [6] and by the vertex detector (VXD) [7], with a combined impact parameter resolution in $r\phi(rz)$ parameterized as $\sigma = 11(38) \oplus 70/p\sqrt{\sin^3\theta}$ $\mu$m, where $p$ is expressed in GeV/c. The liquid argon calorimeter [8] was used for triggering, thrust axis determination, and jet finding. The $\langle$ rms $\rangle_{xyz}$ profile of the SLC beams was approximately $2.4 \times 0.8 \times 700$ $\mu m^3$ at the interaction point (IP). The x and y positions of the IP were continuously measured, using reconstructed tracks from $\sim 30$ sequential hadronic $Z^0$ decays, giving $\sigma_{xy}^{IP} = 7 \pm 2$ $\mu$m [9]. The z position was measured on an event-by-event basis using the median z position of tracks at their point-of-closest-approach to the IP in the xy plane, with a resolution of $\sigma_z \approx 52$ $\mu$m for $Z^0 \to b\bar{b}$ events [9].

A detailed simulation of the detector and physics processes was used in this analysis. Hadronic $Z^0$ decays were generated using JETSET 6.3 [10] adjusted to reproduce data from other $e^+e^-$ experiments. The fragmentation function for $b$- and $c$-quarks followed the Peterson parameterization [11] with $\epsilon_b = 0.006$ and $\epsilon_c = 0.06$, respectively. A detailed description of the $B$-hadron decay model may be found in Ref. [9]. Beam related backgrounds and detector noise were simulated by overlaying random trigger events that occurred in close time proximity to a $Z^0$ decay. Detector response was simulated with GEANT 3.15 [12].

Hadronic $Z^0$ decays were selected by requiring at least seven reconstructed tracks, total track energy greater than 18 GeV, and $|\cos\theta_{thrust}| < 0.71$ (thrust axis within CDC-VXD acceptance). A sample of 29,400 events was selected with a nonhadronic background estimated to be $< 0.1\%$.

A set of *quality* tracks for use in heavy quark tagging and vertexing was selected. Tracks measured in the CDC were required to have $\geq 40$ hits have a hit at a radius r $<$ 39 cm, have a transverse momentum $p_{xy} > 0.4$ GeV/c, extrapolate to the IP within 1 cm in xy and 1.5 cm in z, and have a good fit quality ($\chi^2/d.o.f. < 5$). Tracks were required to have at least one associated hit in the VXD, and a combined CDC/VXD fit with $\chi^2/d.o.f. < 5$. Tracks with a 2-D ($r\phi$) impact parameter $\delta > 3$ mm or a 2-D impact parameter error $\sigma_\delta > 250$ $\mu$m were removed. Tracks from identified $\gamma$ conversions, $K^0$ or $\Lambda^0$ decays, were also removed. A discrepancy in the fraction of tracks passing quality track selection between data and MC was corrected by applying a momentum and angle dependent correction of $\sim 6\%$ to the MC [9].

A sample of 4299 events with an enriched $Z^0 \rightarrow b\bar{b}$ content was tagged by choosing hadronic events that have three or more quality tracks with normalized 2-D impact parameter $\delta/\sigma_\delta > 3.0$. From the MC, this algorithm was determined to be 60% efficient at identifying $Z^0 \rightarrow b\bar{b}$ events providing a 90% pure sample for a sample with $\tau_B = 1.510$ ps. This tag had a minimal effect on the measured lifetime due to a subsequent cut on flight distance, as will be discussed later.

Secondary vertex reconstruction and selection proceeded in three stages. In the first stage, quality tracks with $p > 1.0$ GeV/c were used to construct secondary vertices. Two-prong secondary vertices were constructed from all pairs of these tracks in an event hemisphere, defined with respect to the thrust axis, if they extrapolated to within three standard deviations of a common point in that hemisphere. If a track was shared between a pair of two-prong vertices, a three-prong vertex was also constructed. Four-prong vertices were constructed in a similar way. Tracks from three- and four-prong vertices were required to extrapolate to within three standard deviations of a common point. The vertex fit probability was required to be $> 5\%$. To reduce the number of vertices containing tracks from the IP, the distance between each vertex and the IP was required to be greater than 1 mm, and at least one track in each vertex was required to have $\delta/\sigma_\delta > 2.5$. As a result, 97.8% of tagged events contained at least one selected vertex.

On average, 6.5 vertices per hemisphere remained at this stage of the analysis. Figure 1(a) shows the number of vertices-per-event passing these selection criteria. The first column of Table I shows the composition of the vertices as determined from the MC. A vertex was classified as *'b'* if all the tracks came from the weak-decay vertex of a $B$ hadron, and as a *'b + cascade c'* if at least one track came from the weak decay of a $B$ hadron and the rest from the weak decay of a hadron containing a cascade $c$-quark. The other categories were defined similarly.

Because of the loose vertexing criteria, most geometric vertices were reconstructed; however, tracks were allowed to contribute to more than one vertex. In the second stage of vertex selection, all possible sets (partitions) of independent vertices in each event hemisphere were found; i.e., no vertices shared tracks. Events were rejected if the total number of partitions exceeded 1000. This removed 1.0% of tagged events. For each hemisphere, we selected the partition with maximum value of the product $M = \prod_{all\ vertices} P(\chi^2, d.o.f.)$, where $P(\chi^2, d.o.f.)$ represents the vertex fit probability and the product includes all vertices in a given partition. The partition selection criterion was chosen to provide a high efficiency for finding vertices with lifetime information and low backgrounds. The true $B$-hadron decay vertex was not necessarily reconstructed. Column 2 of Table I shows the vertex composition after selection of the best partition, and Figure 1(b) displays the distribution of the product $M$. To further reduce the background, particularly from the IP, the angle between the vertex line-of-flight and the nearest jet axis was required to be less than 150 mrad, and the vertex tracks were required to have transverse momentum with respect to the vertex line-of-flight greater than 0.07 GeV/c. A sample of 5856 vertices (0.69 vertices/hemisphere on average) remained at this stage of the analysis. Figure 1(c) shows the distribution of the number of prongs per vertex.

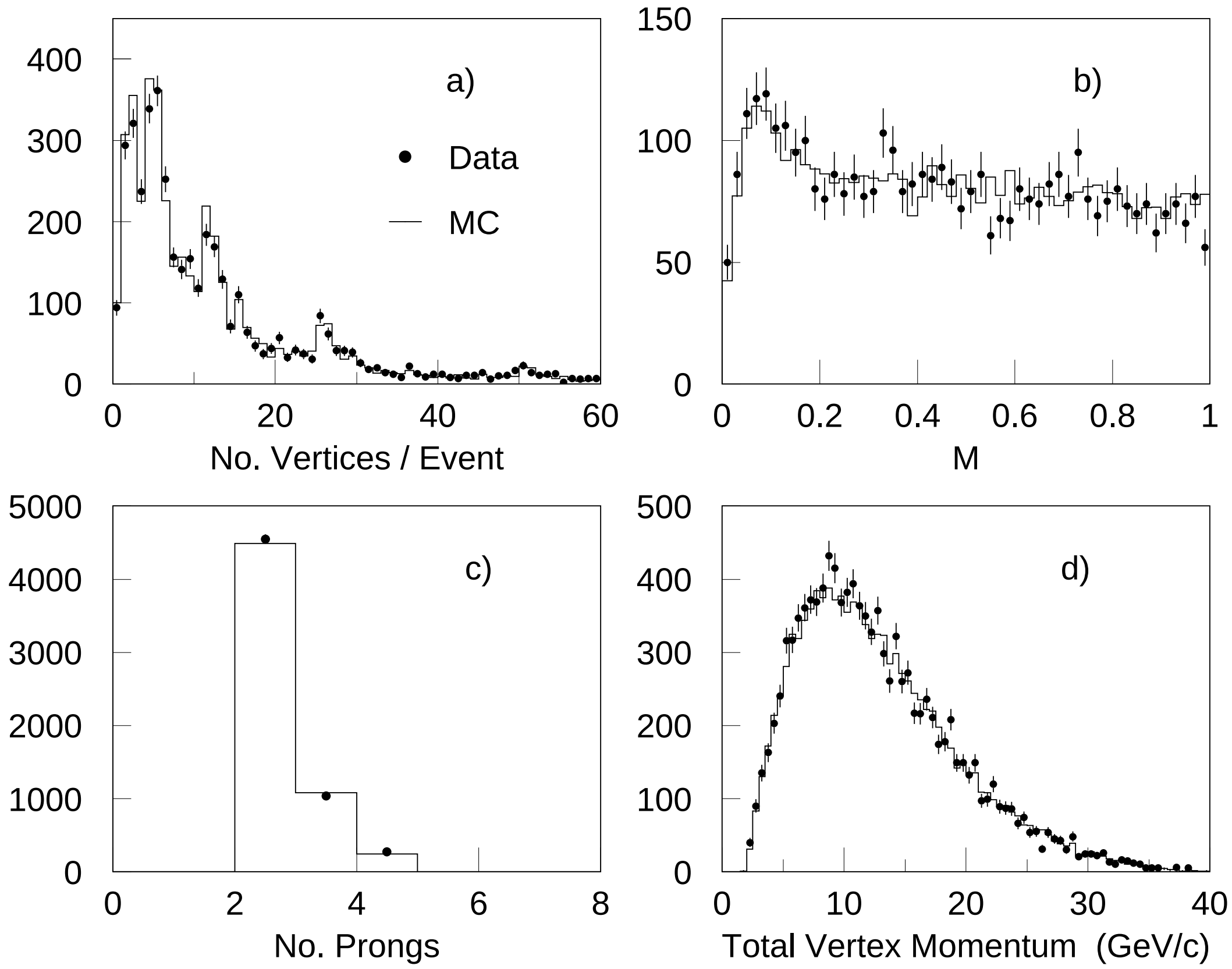

Figure 1. Distribution of (a) number of vertices per event, (b) $M$, (c) number of prongs per vertex, (d) total vertex momentum for data (points) and MC (histogram) with $\tau_B = 1.510$ ps, at different stages of vertex selection.

TABLE I. Vertex type in sample according to Monte Carlo.

| Vertex type | Stage 1: after initial vertex selection (%) | Stage 2: after partition selection (%) | Stage 3: final sample (%) |
|---|---|---|---|
| $b$ | 19 | 21 | 23 |
| Cascade $c$ | 14 | 24 | 26 |
| $b$ + cascade $c$ | 48 | 35 | 37 |
| $b$ + other | 9 | 5 | 3 |
| Cascade $c$ + other | 5 | 6 | 3 |
| Primary $c$ | 3 | 5 | 6 |
| IP | 0.5 | 1.3 | 0.4 |
| Other | 2 | 3 | 2 |
| Vertex multiplicity per hemisphere | 6.5 | 0.7 | 0.6 |

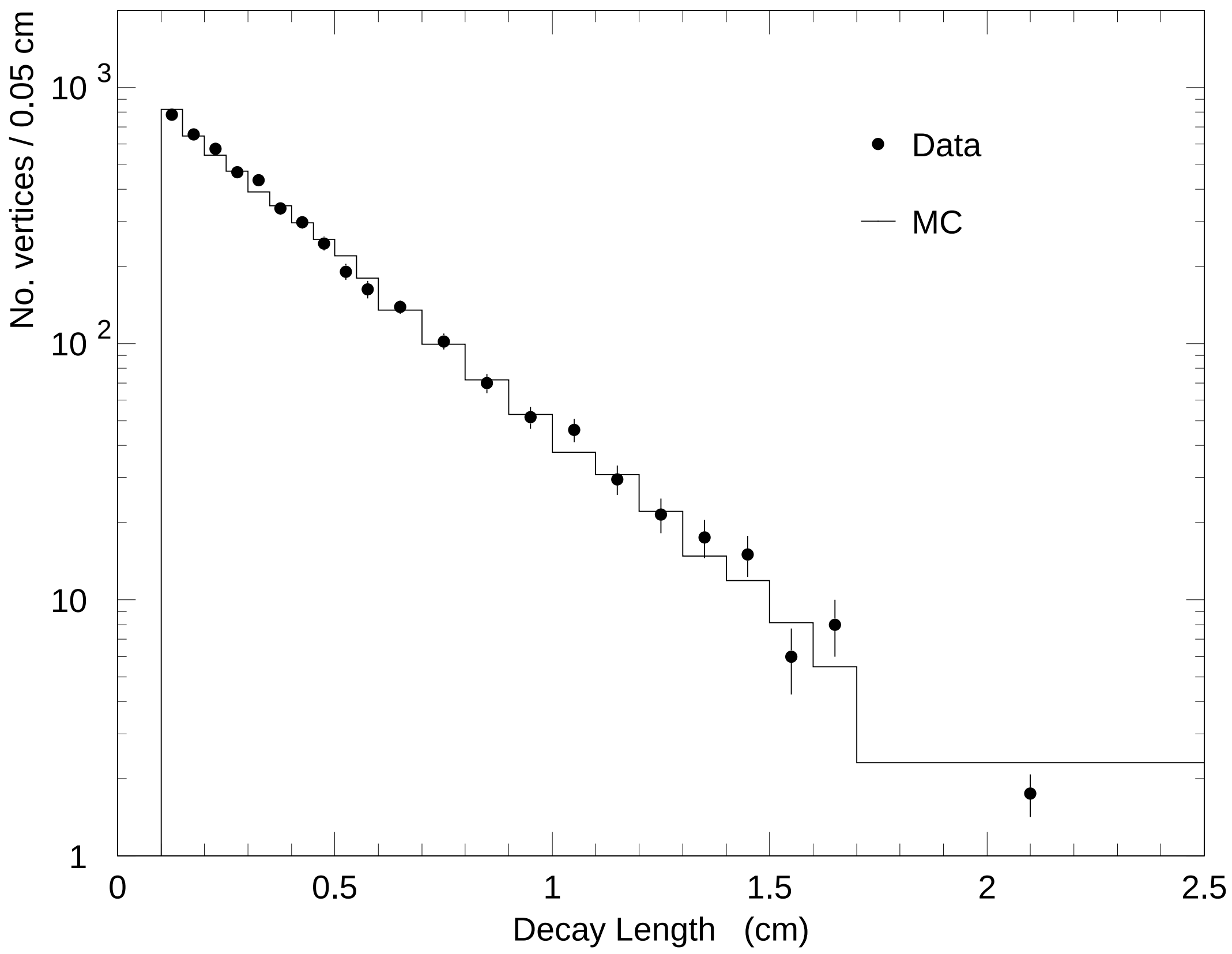

Figure 2. Decay length distribution for vertices passing all selection criteria in data (points) and MC (histogram). The MC distribution corresponds to that with the best-fit lifetime.

In the third stage of vertex selection, one vertex was selected in each hemisphere in order to avoid multiple counting. For hemispheres containing more than one vertex, the vertex closest to the IP was selected to enhance contributions from $b$ vertices over cascade $c$ and other vertices. As a result, 60% of hemispheres in the 4299 tagged events contained a selected vertex. Of these, 92% had a selected vertex containing tracks associated with the weak decay of the $B$ hadron. Figure 1(d) displays the total vertex momentum $|\sum_i \vec{p_i}|$ distribution at this stage. Column 3 of Table I, shows the vertex composition in the final sample. Since this analysis is sensitive to the accuracy of the MC modeling, we have verified that data and MC matched closely at the various stages of vertex selection (see Fig. 1).

The $B$-hadron lifetime was extracted from the decay length distribution, i.e., the distance between secondary vertices and the IP (see Fig. 2). A $\tau_B$ dependence was introduced into the MC by reweighting the decay length distribution of the $B$-hadron component. The MC contained two sets of $Z^0 \rightarrow b\bar{b}$ decays. In the first set, $B$ mesons (baryons), representing 91.1% (8.9%) of all $B$ hadrons, were generated with an average lifetime of 1.55 ps (1.10 ps), and the average lifetime for $B$ mesons (baryons) was 2.00 ps (1.42 ps) in the second set. The average value of $\tau_B$ in the two MC sets was thus 1.510 ps and 1.948 ps, respectively. A binned likelihood was computed for average lifetime values ranging between 0.7 and 2.3 ps; bins in the decay length distribution were combined to have a minimum of ten entries. Using all MC samples, the maximum-likelihood method yielded an average $B$-hadron lifetime of $\tau_B = 1.564 \pm 0.030$ ps with a $\chi^2/d.o.f. \ = \ 27.9/20$.

Systematic checks of the method and the detector modeling were performed. Independent samples of the $\tau_B = 1.510$ ps MC were used as data yielding lifetimes in agreement with the generated lifetime. The lifetime was also extracted from the data using the $\tau_B = 1.510$ ps and $\tau_B = 1.948$ ps MC $Z^0 \rightarrow b\bar{b}$ samples separately, yielding values consistent with the reported result. In the two MC sets, the average values of the decay length distributions for each vertex type at the same reweighted $\tau_B$ were found to agree to within 1%. Variations in the measured lifetime between different epochs of the run, i.e., under different run conditions, were found to be within the statistical error of our quoted value. No systematic dependence on azimuthal angle could be inferred from the lifetime measurements from four different azimuthal sections of the detector. We verified that the initial heavy-quark tag had negligible effect by comparing the vertex composition and decay length distributions for tagged and untagged $Z^0 \rightarrow b\bar{b}$ events that passed all other cuts in the analysis. The 1 mm decay length cut reduced the backgrounds, especially from $Z^0 \rightarrow c\bar{c}$ events, and was found to reduce the sensitivity to the choice of tagging criteria. The tag was also investigated by considering only hemispheres opposite jets which were tagged by requiring at least 3 tracks in the jet with $\delta/\sigma_\delta > 3.0$. This yielded an average lifetime of $\tau_B = 1.616 \pm 0.052$ ps, which was independent of the heavy-quark tag and consistent with the reported result. We verified that the choice of vertex in the last stage of the analysis had little effect on the lifetime measurement.

The systematic error due to detector modeling (see Table II) was dominated by the modeling of the reconstruction of quality tracks in the detector. This effect included an overall normalization between data and MC after quality track selection, as well as a 0.3 track variation in quality track multiplicity between different run periods [9]. Track parameters were varied to reflect uncertainties in the radial and longitudinal alignment within the VXD and to account for the differences between data and MC in the tails of the IP position distribution along the z-axis. The IP x and y positions were Gaussian-smeared by 100 $\mu$m for 0.25% of the MC events to reproduce tails observed in the data.

The systematic errors due to the heavy-quark physics modeling are also shown in Table II. Systematic errors due to $b$ and $c$ fragmentation were determined by varying $\epsilon_b$ and $\epsilon_c$ in the Peterson function used in our MC [11], according to $\langle x_E \rangle = 0.700 \pm 0.011$ for $b$ fragmentation and $\langle x_E \rangle = 0.49 \pm 0.03$ for $c$ fragmentation [13]. The $b$-fragmentation systematic error also includes a contribution due to the uncertainty in the shape of the fragmentation function, which was estimated by using the Bowler-Lund parameterization [14]. Sensitivity to the charm content of $B$-hadron decays was estimated by taking into account the uncertainty in the $B^0$ and $B^+$ branching ratios [3]. The uncertainties were inflated by 100% to account for the fact that no data are available for $B_s$ and $B$ baryon branching ratios. The absolute fraction of $B$ baryons in $Z^0 \to b\bar{b}$ events was varied by $\pm 0.05$. The absolute fractions of $D^+$ and charm baryons in $Z^0 \to c\bar{c}$ events were varied by $\pm 0.037$ [15] and $\pm 0.03$ respectively. $R_b$ and $R_c$ [16] were varied by the uncertainty in the world average [3]. The multiplicity of tracks from $B$-hadron decays was varied by $\pm 0.3$ tracks to reflect the current uncertainty in this value [17]. The MC momentum spectra of $D^+$ and $D^0$ mesons from $B$-hadron decays were adjusted to agree with recent CLEO data [18]. The lifetimes of $D^+$, $D^0$, $D_s$, and $\Lambda_c$ were varied by the uncertainty in their world averages [3].

TABLE II. Summary of systematic errors.

| Error source | $\Delta\tau_B$ (ps) |
|---|---|
| Quality track multiplicity | 0.013 |
| VXD alignment | 0.005 |
| $\mathrm{IP}_{xy}$ position tails | <0.001 |
| $b$ fragmentation | 0.018 |
| $c$ fragmentation | 0.005 |
| Charm content of $B$ decays | 0.014 |
| $B$-baryon fraction | 0.004 |
| Charm-hadron fraction | 0.012 |
| $R_b$ | <0.001 |
| $R_c$ | 0.002 |
| $B$-hadron decay multiplicity | 0.009 |
| $D$-momentum spectrum in $B$ decays | 0.010 |
| Charm-hadron lifetime | 0.004 |
| Minimum decay length in lifetime measurement | 0.006 |
| Decay-length binning | 0.007 |
| Decay-length cut in vertex selection | 0.010 |
| Vertex and partition limits | 0.003 |
| Monte Carlo statistics | 0.008 |
| Total systematic errors | 0.037 |

Decay-length bin size was varied from 0.25 mm to 2 mm, and the minimum decay length used to extract the lifetime was varied from 1 mm to 3 mm to determine the variation in the measured lifetime. Samples passing different decay-length cuts at the vertex reconstruction stage were fitted over the same range to assess the effect of this cut. The systematic error due to the cut on the maximum number of partitions was estimated to be negligible. This was accomplished by weighing the $B$-hadron decay-length distribution in the Monte Carlo final sample to reproduce the distribution prior to the cut.

In conclusion, an inclusive vertexing technique was used to determine the average $B$-hadron lifetime from a sample of 50 k $Z^0$ decays. The measured lifetime was $1.564 \pm 0.030\,(stat) \pm 0.037\,(syst)$ ps, consistent with recent results from other experiments [4,5].

We would like to thank the personnel of the SLAC accelerator department and the technical staffs of our collaborating institutions for their outstanding efforts.